\documentclass[12pt]{extarticle}
\usepackage{a4wide}
\usepackage{graphicx}
\usepackage{amssymb}
\usepackage{amsmath}
\usepackage{gensymb}
\newcommand{\be}{\begin{equation}}
\newcommand{\ee}{\end{equation}}
\newcommand{\ba}{\begin{eqnarray}}
\newcommand{\ea}{\end{eqnarray}}

\begin{document}

\begin{titlepage}
\begin{flushright}
\end{flushright}
\vfill
\begin{center}
{\Large\bf Fermionic dark matter with pseudo-scalar Yukawa interaction}
\vfill
{\bf Karim Ghorbani\footnote{kghorbani@ipm.ir}}\\[1cm]
{Physics Department, Faculty of Sciences, Arak University, Arak 38156-8-8349, Iran}
\end{center}
\vfill
\begin{abstract}
We consider a renormalizable extension of the standard model whose fermionic 
dark matter (DM) candidate interacts with a real singlet pseudo-scalar via a pseudo-scalar 
Yukawa term while we assume that the full Lagrangian is CP-conserved in the classical level.
When the pseudo-scalar boson develops a non-zero vacuum expectation value, spontaneous 
CP-violation occurs and this provides a CP-violated interaction of the dark sector 
with the SM particles through mixing between the Higgs-like boson and the SM-like Higgs boson.
This scenario suggests a minimal number of free parameters.      
Focusing mainly on the indirect detection observables, we calculate the dark matter 
annihilation cross section and then compute the 
DM relic density in the range up to $m_{\text{DM}} = 300$ GeV. 
We then find viable regions in the parameter space constrained by 
the observed DM relic abundance 
as well as invisible Higgs decay width in the light of 125 GeV 
Higgs discovery at the LHC.    
We find that within the constrained region of the parameter space, there exists
a model with dark matter mass $m_{\text{DM}} \sim 38$ GeV annihilating predominantly 
into $b$ quarks, which can explain the Fermi-LAT galactic gamma-ray excess.

\end{abstract}
\vfill
keywords: Beyond the standard model, dark matter experiments, dark matter theory
\vfill
{\footnotesize\noindent }

\end{titlepage}

\section{Introduction}
\label{int}
The search for deciphering the identity of the dark matter (DM) has been  
under intense scrutiny since long ago, see reviews in \cite{Bertone:review,Bergstrom:review}. 
There is a strong confidence that about 25 per cent of the matter 
content of the Universe is made of a new very long lived particle or 
particles, the so called dark matter \cite{Ade:Planck}. 
The search for DM signals divides into {\it direct} detection and {\it indirect} 
detection methods. The former approach relies on the 
DM scattering with ordinary matter while the latter avenue depends on the 
dark matter annihilation processes.    

In case there is a sensible interaction of DM with ordinary matter, in 
direct detection approach, the experiment is set up so as to measure
the recoil energy of the nuclei induced by DM scattering off nucleons \cite{Goodman:1984dc,Drukier:1986tm}. 
In this regards, the first results of LUX dark matter experiment \cite{Akerib:LUX} 
although finds no evidence for DM interaction it provides us with impressive bounds
on the DM-nucleon scattering cross section in a wide range of DM mass.    
Along the same line, dark matter results from XENO100 experiment \cite{Aprile:XENON100}  
again finds low spin-independent DM-nucleon scattering rate.     

On the other hand, within the indirect detection method, lies 
the accurate measurement of the dark matter density. 
The Planck experiment recently obtained the cold dark matter (CDM) density 
based on high precision measurement of the acoustic peaks in the 
cosmic microwave background \cite{Ade:Planck}
\ba
\nonumber
\Omega_{\text{CDM}} h^2 = 0.1196\pm 0.0031\,.  ~~~~~~(68\%~~ \text{C.L.}) \,.
\ea   
In agreement with the Planck result, WMAP temperature and polarization data including
low multipoles \cite{Hinshaw:2012aka} provided us with the cold dark matter density as
\ba
\nonumber
\Omega_{\text{CDM}} h^2 = 0.1138\pm 0.0045\,.  ~~~~~~(68\%~~ \text{C.L.}) \,.
\ea  
In addition, as indirect detection, dark matter pair annihilation can produce 
potentially measurable anomalous gamma-rays, cosmic rays and also neutrinos.
Gamma-rays are particularly interesting and may be observable by 
the {\it Fermi} Gamma-Ray Space Telescope \cite{Baltz:2008wd}. 
A promising place on the sky to look for the gamma-rays are the
central region of the Milky Way which contains a high density of DM
and it is relatively close to us, see discussions in \cite{Goodenough:2009,Hooper:2010mq,Boyarsky:2010dr}.  

Besides the possibility of direct and indirect detection of DM, it is plausible to have DM 
production at particle colliders \cite{Beltran:2010ww,Goodman:2010ku,Bai:2010hh}.

Motivated by the observational developments discussed above, the question now is 
about the nature of the dark matter.
So let us take a look at the theoretical undertakings. As it is well-known the Standard Model 
(SM) of particle physics is lacking a proper candidate for DM. 
A copious number of theories beyond the SM exist which propose some kind of
dark particle candidates generically called weakly interacting 
massive particles (WIMPs) in order to explain the observed 
relic density. 

The most vastly investigated DM candidate as a WIMP is the lightest 
supersymmetric particle (LSP) which is a stable particle 
in supersymmetric (SUSY) models with conserved R-parity, see \cite{Jungman:1995df} 
and references therein. However, being well motivated theoretically, 
the presence of large number of free parameters have hindered our 
predictivity within SUSY models. On the contrary, in models with universal
extra dimensions (UED), the size of the extra dimension is the only
parameter of the model which dominates the physics \cite{Cheng:2002ej}. 
The lightest Kaluza-Klein particle whose mass is the inverse of 
the compactification radius, is the DM candidate which 
is stable due to the conserved Kaluza-Klein parity.      
Relying on a new symmetry at the TeV scale dubbed T-parity, 
in the little higgs model introduced in \cite{Cheng:2003ju} 
to cure the little hierarchy problem, emerges the lightest T-odd 
new particle which may serve as a WIMP. 

The minimal extension of the SM is the addition of a gauge singlet real 
scalar field \cite{Silveira:1985rk} or a gauge singlet complex 
scalar field \cite{McDonald:1993ex} to the SM with $Z_{2}$-parity imposed 
on the new fields to ensure their stability as 
DM candidates. A minimal dark matter model is also introduced in \cite{Cirelli:2005uq}
in which the new fermionic or bosonic field has only gauge interaction. 
Moreover, a minimal extension of the SM is constructed by the inclusion 
of a hidden sector incorporating
a gauge singlet scalar field and a gauge singlet fermionic field \cite{Kim:2008pp}. 
The fermionic field interacts with the SM fields only through the singlet 
scalar field while the latter has triplet and quartic scalar interactions with 
the SM higgs doublet. Since the new fermion is assumed to be charged under 
a global unbroken U(1) symmetry while the SM fields are neutral 
under the same symmetry, there is no direct interaction between 
the singlet fermion and SM particles. In this model singlet fermion 
is the DM candidate.            

Recently, motivated by the null result from direct detection of DM, 
it has been thought that the WIMP dark matter may interact in such a  
way as to leave a trace in one experiment but not necessarily in another one.
One example in this regards, is a model put forward in \cite{Boehm:2014hva} suggesting a new 
type of dark matter dubbed, coy dark matter. In this model it is shown that the  
proposed dark matter can explain the observed extended gamma-ray flux originating 
from the galactic center without expecting any signals from direct detection or elsewhere.    

Additionally, in \cite{Pospelov:2011-CPfermionic,LopezHonorez:2012-CPfermionic} the possibility 
of having no signature from direct detection is studied with singlet fermionic dark matter 
whose interaction with the SM particles is through a Higgs portal. 
Using a super-renormalizable Higgs portal in \cite{Pospelov:2011-CPfermionic}, a dominant
CP-violated coupling for DM is found which leads to a suppression of order $\sim 10^{3}$
for all visible Higgs decay channels. However, no suppression of this magnitude is 
found at the LHC \cite{Barger:Higgs-Width}. On the other hand, within the framework of 
effective field theory in \cite{LopezHonorez:2012-CPfermionic} a 5-dimensional 
Higgs portal is considered with CP-conserved and CP-violated components. It is shown
in \cite{LopezHonorez:2012-CPfermionic} that both components are necessary in order for
DM with $60$ GeV $\lesssim m_{\text{DM}} \lesssim 2$ TeV to be in 
consistent with direct detection results. Along the same line, 
in \cite{Esch2013-GeneralCase} an updated analysis is performed for a general
case including both CP-odd and CP-even interactions within a UV completion 
model introduced in \cite{LopezHonorez:2012-CPfermionic}.        

In this article we consider a minimal extension of the SM in which a fermionic 
dark matter interacts with a pseudo-scalar mediator via a pseudo-scalar 
Yukawa interaction. We assume our Lagrangian to be CP-conserved in the 
classical level and therefore the pseudo-scalar mediator is allowed to interact 
with DM only through $g \phi \chi \gamma^{5} \chi$. When pseudo-scalar boson develops 
a non-zero vacuum expectation value, the CP symmetry is broken spontaneously and on
the other hand, the Higgs portal $\phi^{2} H^{\dagger}H$ provides 
a link between DM sector and the SM particles.        
As a new development we focus in this work on the connection between the invisible Higgs
decay width measurements and indirect detection observations. 
It is known a priori that direct detection of DM is suppressed in our scenario. 
Our prime motivation here is to firstly find the viable parameter space constrained by 
DM relic density observations and if applicable by the invisible Higgs decay 
width and secondly we quest for a parameter region in the constrained 
viable parameter space to explain the Galactic Center anomalous gamma ray excess.

The structure of the paper is as follows.
In Section.~(\ref{model}) a fermionic dark matter model with only CP-odd coupling for DM 
is considered and relations among the relevant couplings are derived. 
In Section.~(\ref{constrain}) Higgs decay width for decays into DM and scalar mediator 
are calculated and constraints from measured Higgs decay width is considered.   
We calculate the DM annihilation cross sections in Section.~(\ref{relic-density}) 
and then compute the relic density for various sets of parameters. 
We also find in this section the viable parameter space constrained by observed relic density and 
measured invisible Higgs decay width. In Section.~(\ref{direct}) we provide a 
detailed derivation for the DM-nucleus elastic scattering for DM with pseudo-scalar coupling. 
We find in Section.~(\ref{gama-ray}) that it is possible to explain the observed gamma ray excess from 
the Galactic Center within our constrained parameter space. 
We finish up with a conclusion in Section.~(\ref{con}).

\section{Singlet fermionic dark matter} 
\label{model}
The model we consider here is a renormalizable extension of the Standard Model (SM) Lagrangian 
with two new extra fields, one Dirac fermion field $\chi$ and one real pseudo-scalar field $\phi$. 
The new fields are SM gauge singlets and the fermionic field is 
charged under a global $\text{U(1)}_{\text{DM}}$ symmetry.
Since all the SM fields are singlet under the global symmetry, the SM particles 
interact with the dark sector only via the Higgs portal. We assume in the following
that our full Lagrangian is CP-invariant in the classical level.  

The model Lagrangian therefore consists of the following parts:
\ba
{\cal L} = {\cal L_{\text{SM}}}+{\cal L}_{\text{Dark}} + {\cal L_{\phi}}+{\cal L}_{\text{int}}\,,
\ea     
where ${\cal L}_{\text{Dark}}$ introduces the singlet Dirac field which does not undergo any 
mixing with the SM fermions due to the presumed global $\text{U(1)}_{\text{DM}}$ symmetry of the 
singlet fermion with   
\ba
{\cal L}_{\text{Dark}} =  \bar{\chi} (i {\not}\partial-m_{D}) \chi  \,,
\ea     
and ${\cal L_{\phi}}$ is a renormalizable Lagrangian for the  pseudo-scalar boson as
\ba
{\cal L_{\phi}}   =  \frac{1}{2} (\partial_{\mu} \phi)^2 - \frac{m^{2}_{0}}{2}\phi^2 -\frac{\lambda}{24}\phi^4  \,.
\ea    
The interaction Lagrangian itself, ${\cal L}_{\text{int}}$, consists of a pseudo-scalar Yukawa 
term and an interaction term incorporating SM-higgs doublet and singlet pseudo-scalar, 
\ba
{\cal L}_{\text{int}}   =  -ig\phi \bar{\chi}\gamma^{5}\chi - \lambda_{1} \phi^2 H^{\dagger}H  \,.
\ea    
It is readily seen that the Lagrangian ${\cal L}$ is CP-invariant since under the parity transformation
$\phi(t,\vec{x}) \to -\phi(t,-\vec{x})$ and $\chi(t,\vec{x}) \to \gamma^{0} \chi(t,-\vec{x})$ 
in which the scalar field $\phi$ carries odd parity and under charge conjugation transformation
we have $\phi(t,\vec{x}) \to \phi(t,\vec{x})$ and 
$\chi(t,\vec{x}) \to \chi(t,\vec{x})$.
The Higgs field, $H$, is a SM $\text{SU(2)}_{\text{L}}$ 
scalar doublet. On the other hand, the SM Higgs potential introduces the quartic self coupling 
of the Higgs field as
\ba
V_{H}  =  -\mu^{2}_{H} H^{\dagger}H - \lambda_{H} (H^{\dagger}H)^2  \,.
\ea 
The Higgs field develops a non-zero vacuum expectation value (vev) which gives rise to 
the electroweak spontaneous symmetry breaking. The fluctuation about the vev is described by 
the scalar field $\tilde h$ such that 
\ba
H = \frac{1}{\sqrt{2}} \left( \begin{array}{c}
                                0  \\
                                v_{H}+\tilde h
                       \end{array} \right)\,,
\ea
where $v_{H}$ = 246 GeV. In addition we assume in this model that the 
pseudo-scalar singlet also acquires a non-zero
vev as
\ba
\phi =  v_{\phi} + S\,.
\ea
This consequently leads to the spontaneously breaking of CP symmetry.    
The global $\text{U(1)}_{\text{DM}}$ symmetry is conserved even after the 
spontaneous symmetry breaking and thus,
this ensures the stability of the fermionic singlet which is a necessary 
condition for a proper dark matter candidate.

\begin{figure}
\begin{minipage}{0.4\textwidth}
\includegraphics[width=\textwidth,angle =-90]{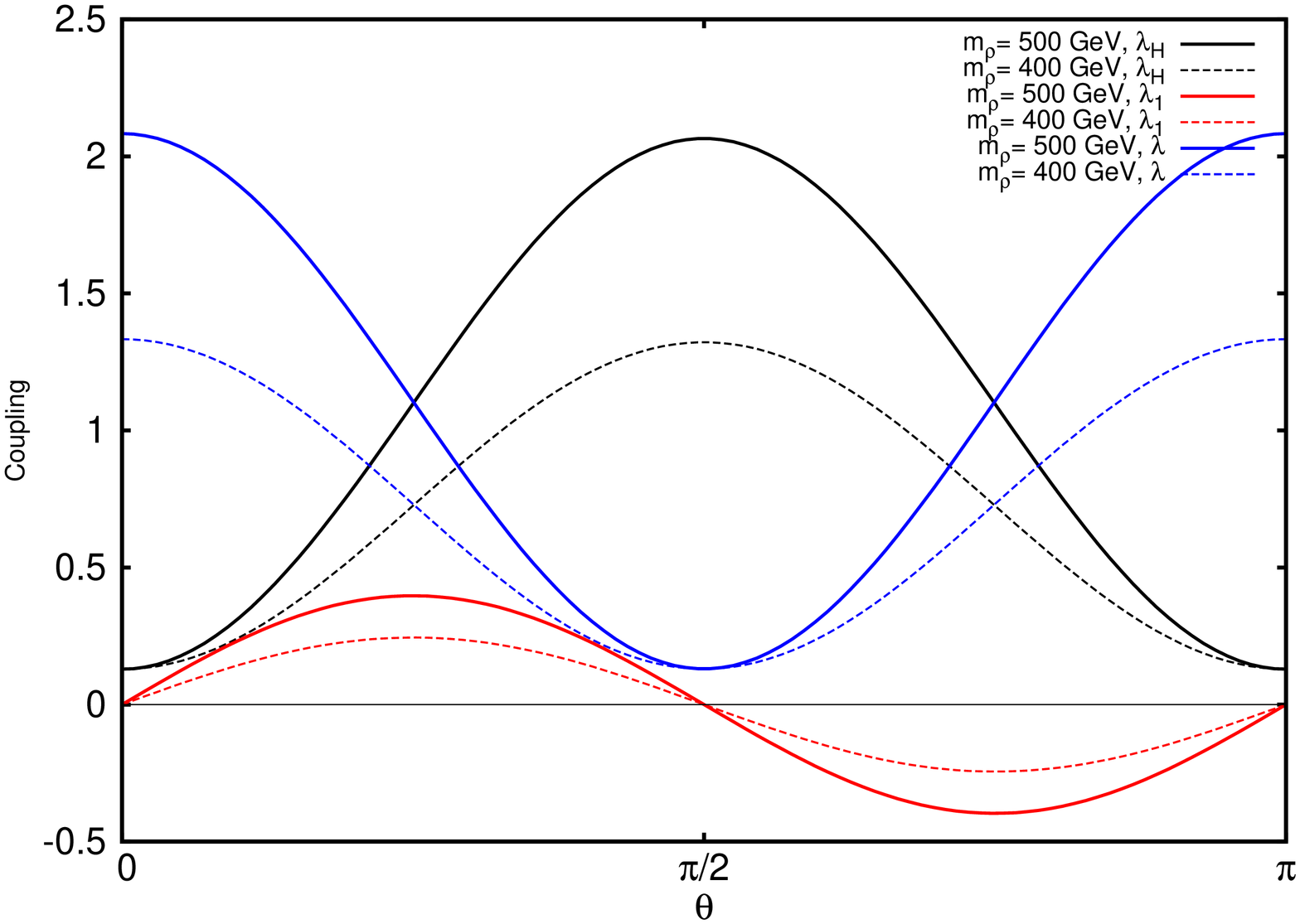}
\end{minipage}
\hspace{1.8cm}
\begin{minipage}{0.4\textwidth}
\includegraphics[width=\textwidth,angle =-90]{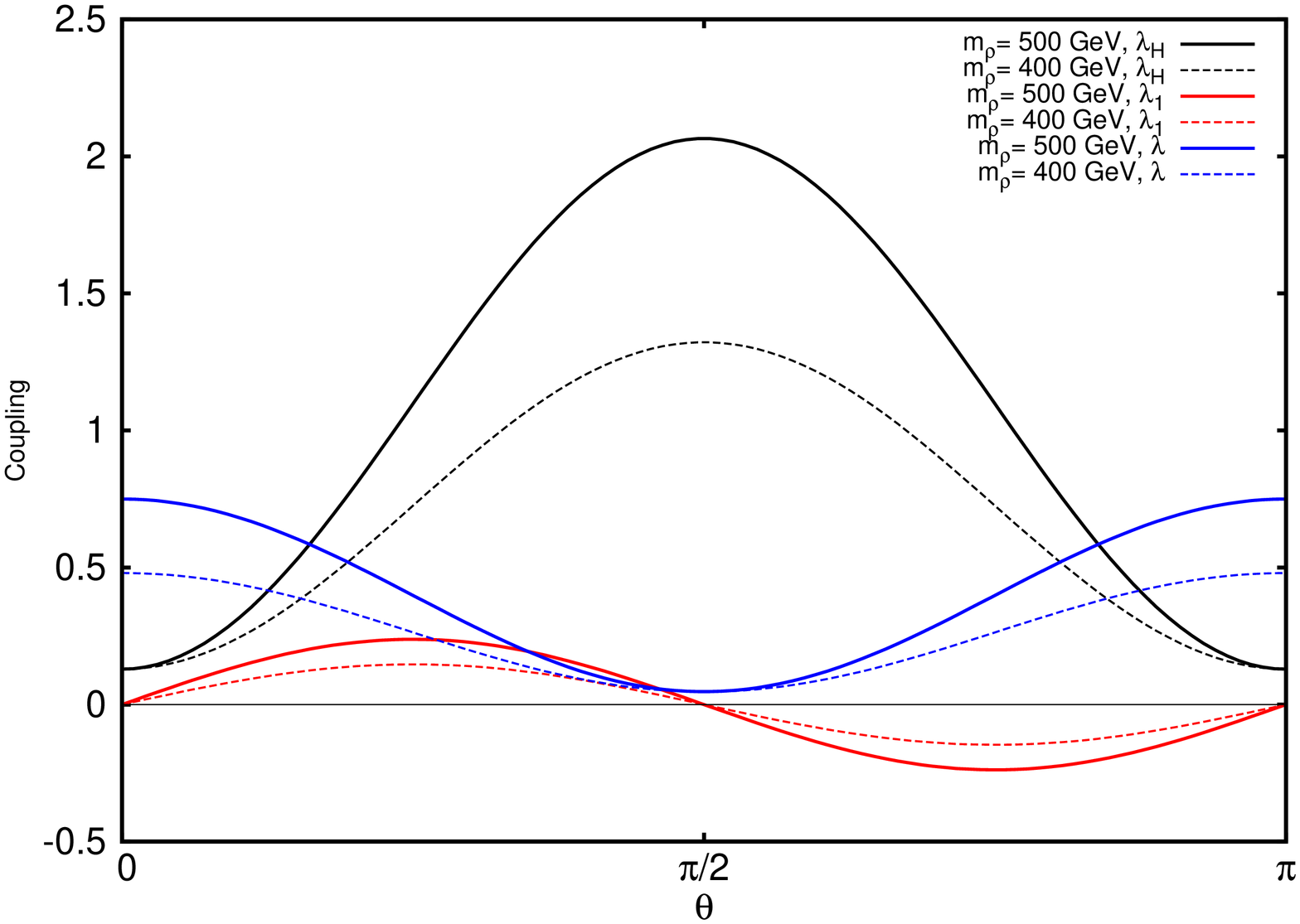}
\end{minipage}
\caption{Shown are the quartic couplings as a function of the mixing angle 
at $v_{H}$ = 246 GeV for two different values of the scalar boson 
mass, $m_{\rho}$ =  400 GeV and 500 GeV. In the left panel $v_{\phi}$ = 600 GeV 
and in the right panel $v_{\phi}$ = 1000 GeV.}
\label{drawcoupling}
\end{figure}

From the minimization condition of the potential, i.e., 
\ba
\frac{\partial V}{\partial H}|_{<H> = v_{H}/\sqrt{2}} = \frac{\partial V}{\partial \phi}|_{<\phi> = v_{\phi}} = 0  \,,  
\ea
we can express two parameters of the model in terms of the vevs and 
quartic coupling by the relations

\ba
\label{mass-reduce}
m^{2}_{0} = -\frac{\lambda}{6} v^{2}_{\phi} - \lambda_{1} v^{2}_{H} \,,
\nonumber \\
\mu^{2}_{H} = -\lambda_{H} v^{2}_{H} - \lambda_{1} v^{2}_{\phi}\,.
\ea
We now turn back to the Lagrangian and pick out entries of the mass matrix 
associated with the SM-higgs field, $\tilde h$, and
the scalar field $S$,
\ba
\label{mass-s}
m^{2}_{S} = \frac{\partial^2 V}{\partial S^2} = \frac{1}{3}\lambda v^{2}_{\phi}\,,
\ea
\ba
\label{mass-h}
m^{2}_{\tilde h} = \frac{\partial^2 V}{\partial h^2} = 2 \lambda_{H} v^{2}\,,
\ea
and
\ba
m^{2}_{\tilde h,S} = \frac{\partial^2 V}{\partial S \partial h} = 2 \lambda_{1} v_{\phi} v\,,
\ea
in which to obtain the relations above we have used Eq.~(\ref{mass-reduce}).
We then indicate the mass eigenstates $h$ and $\rho$ as following 
by defining the mass mixing angle $\theta$, 
\ba
h = \sin \theta~S + \cos \theta~\tilde h\,, 
\nonumber\\
\rho = \cos \theta~ S - \sin \theta~ \tilde h\,, 
\ea
where,
\ba
\tan \theta = \frac{y}{1+\sqrt{1+y^2}} \,, ~~~ \text{with}~~ y= \frac{2m^{2}_{\tilde h,S}}{m^{2}_{\tilde h}-m^{2}_{S}}\,.
\ea
The two neutral Higgs-like scalars $h$ and $\rho$ given as admixtures of
SM higgs $\tilde h$ and scalar S, have reduced couplings to the SM particles by a factor
$\sin \theta$ or $\cos \theta$. 
The corresponding mass eigenvalues are given by 
\ba
\label{eigenvalues}
m^{2}_{h,\rho} = \frac{m^{2}_{\tilde h}+m^{2}_{S}}{2}\pm \frac{m^{2}_{\tilde h}-m^{2}_{S}}{2} \sqrt{1+y^2}\,, 
\ea
where the upper sign (lower sign) corresponds to $m_{h} (m_{\rho})$. 
In the following we assume that $h$ is the eigenstate of the SM higgs 
with $m_{h}$ = 125 GeV and $\rho$ corresponds to the eigenstate of the singlet scalar. 
It is possible to obtain the quartic couplings in terms of higgs masses, vevs and mixing angle  
\ba
\label{couplings}
\lambda_{H}  = \frac{m^{2}_{\rho} \sin^2 \theta +m^{2}_{h} \cos^2 \theta }{2v^{2}_{H}}\,,
\nonumber\\
\lambda  = \frac{m^{2}_{\rho} \cos^2 \theta +m^{2}_{h} \sin^2 \theta }{v^{2}_{\phi}/3}\,,
\nonumber\\
\lambda_{1} = \frac{m^{2}_{\rho}-m^{2}_{h}}{4v_{H}v_{\phi}} \sin 2\theta.
\ea  
The stability of the potential puts constrains on the quartic couplings as 
$\lambda > 0$, $\lambda_{H} > 0$ and $\lambda \lambda_{H} > 6 \lambda^{2}_{1}$.
One more restriction on the couplings comes from the perturbativity requirement of 
the model which demands $|\lambda_{i}| < 4\pi$.
The set of independent free parameters in the model are considered to be $m_{\chi}$, $m_{\rho}$, 
$g$, $\theta$ and $v_{\phi}$.    
We use the relations given in Eq.~(\ref{couplings}) to display in Fig.~(\ref{drawcoupling}) 
the dependency of the couplings on the mixing angle, $m_{\rho}$ and $v_{\phi}$. 
Two different values are chosen for the scalar boson mass, $m_{\rho}$ = 400 
and 500 GeV while for both cases we take $v_{H}$ = 246 GeV. 
Comparison between our results in the left and right panels are made for two 
different values of $v_{\phi}$, namely 600 GeV and 1000 GeV.    
We find out that both conditions, $\lambda \lambda_{H} > 6 \lambda^{2}_{1}$ and 
$|\lambda_{i}| < 4\pi$ are well fulfilled for the above parameter set. 

\section{Constraint from invisible Higgs decay}
\label{constrain}
Within the Standard Model the total decay width of the Higgs boson is 
$\Gamma^{\text{SM}}_{\text{Higgs}} \approx 4$ MeV \cite{Denner:Higgs-DecayWidth} 
for a Higgs mass of 125 GeV.
For light dark matter mass such that $m_{\chi} < \frac{m_{h}}{2}$ and additionally 
having the condition $m_{\rho} > m_{h}/2$, there is only one new channel open for
the Higgs decay which is kinematically allowed, 
\ba
\label{dark-decay}
\Gamma(h \to \bar \chi \chi ) = \frac{g^2\sin^2\theta}{8\pi} (m^{2}_{h}-4 m^{2}_{\chi})^{1/2}.
\ea
We therefore expect the modification of the total decay width of the Higgs boson as 
\ba
\label{total-width}
\Gamma^{\text{tot}}_{h} = \cos^2\theta~\Gamma^{\text{SM}}_{\text{Higgs}} + \Gamma(h \to \bar \chi \chi). 
\ea 
The invisible branching ratio of the Higgs decay for various channels are investigated recently 
in the light of 125 GeV Higgs discovery at the LHC in \cite{Bai:invisible,Ghosh:invisible,Belanger:invisible}.   
In \cite{Belanger:invisible} a conservative experimental upper limit for the invisible branching fraction 
of the Higgs boson is achieved, $B_{\text{inv}} \lesssim 0.35$. Thus we can obtain 
from Eq.~(\ref{total-width}) an upper limit for the invisible Higgs decay width, 
\ba
\Gamma(h \to \bar \chi \chi ) < \frac{B_{\text{inv}}}{1-B_{\text{inv}}} 
\cos^2\theta~\Gamma^{\text{SM}}_{\text{Higgs}}. 
\ea   
Given the decay width of the Higgs boson into two dark matter particles 
in Eq.~(\ref{dark-decay}) we derive an upper limit for the product $|g\tan \theta|$ as
\ba
|g\tan \theta| < \frac{7.35~(MeV)^{1/2}}{(m^2_{h} - 4 m^{2}_{\psi})^{1/4}}.
\ea
In case we consider light scalar boson with $m_{\rho} < m_{h}/2$, there is one more 
possible decay channel for SM-higgs decay with 
\ba
\label{higgs-decay-ro}
\Gamma(h \to \rho \rho ) = \frac{c^2}{8\pi m_{h}} (1-4 m^{2}_{\rho}/m^{2}_{h})^{1/2}.
\ea
in which 
\ba
c = (4 \sin \theta-6 \sin^3 \theta) \lambda_{1} v_{\phi} 
+ (6 \cos \theta \sin^2 \theta - 2 \cos \theta) \lambda_{1} v_{H} \\
\nonumber
-(\cos^2 \theta \sin \theta) \lambda v_{\phi} - 6 \cos \theta \sin^2 \theta \lambda_{H} v_{H}.
\ea
Our numerical examination shows that the effect of the decay $h \to \rho \rho$ on 
the upper bound of $|g\tan \theta|$ is essentially negligible.

\section{Dark Matter Relic Density}
\label{relic-density}
\begin{figure}
\begin{center}
\includegraphics[width=.72\textwidth,angle =0]{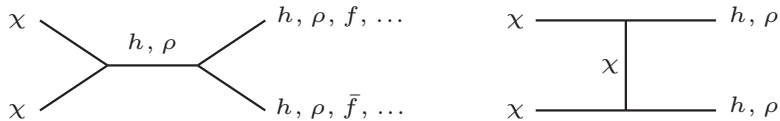}
\end{center}
\caption{Shown are the relevant Feynman diagrams for DM annihilation.}
\label{feynman}
\end{figure}
The problem of dark matter is an interesting instance of freeze-out in the early 
Universe. It is in fact the question of what happens when dark particles ($\chi$) 
go out of equilibrium. 
The pair annihilation of dark particles into pairs of 
SM particles ($\bar{\chi} \chi \to \bar{X} X$) and the inverse processes play a central 
role in our treatment based on the Boltzmann transport equation 
in an expanding Universe. The reason relies on the fact that only the annihilation and 
production reactions can change the number of dark particles in the comoving volume.    

In thermal equilibrium, annihilation of dark particles take place with the same rate 
as their creation processes occur.  
However, an expanding Universe cools down and reaches a point (T$ \ll \text{m}_{\text{DM}}$) 
in which the dark particle interactions freeze out. In fact at the freeze-out temperature 
the annihilation rate of dark particles drops below the Hubble expansion rate.
On the other hand, at temperature T$ \ll \text{m}_{\text{DM}}$, dark particle production reactions
are Boltzmann suppressed since only a small portion of $\bar X X$ have enough kinetic 
energy to produce a $\bar \chi \chi$ pair.  
After freeze-out, the number density, $n_{\chi}$ does not change with time asymptotically. 
We can thus determine the present value of the relic density by solving numerically 
the evolution equation. 

Taking into account the considerations sketched above, the time 
evolution of the number density of the singlet dark matter in 
departure from equilibrium is governed by the Boltzmann equation as
\ba
\frac{dn_{\chi}}{dt} +3Hn_{\chi} = - \langle \sigma_{\text{ann}}v_{\text{rel}} \rangle [n^{2}_{\chi}-(n^{\text{EQ}}_{\chi})^2 ].
\ea
The second term in the left-hand side is the dilution due to the expanding Universe, where $H$
is the Hubble parameter. In the expression $\langle \sigma_{\text{ann}}v_{\text{rel}} \rangle$ thermal averaging is 
understood because particles annihilate with random thermal velocities and directions. The thermal
average of the annihilation cross section times the relative velocity at temperature $T$ 
is obtained by integration over the center of mass energy $\sqrt{s}$ as  
\ba 
\langle \sigma_{\text{ann}} v_{\text{rel}} \rangle = \frac{1}{8 m_{\chi}^4TK^{2}_{2}(\frac{m_{\chi}}{T})}
\int^{\infty}_{4m^{2}_{\chi}} ds~(s-4m^{2}_{\chi})\sqrt{s}~K_{1} (\frac{\sqrt{s}}{T})~\sigma_{\text{ann}}(s)\,, 
\ea   
in which $K_{1,2}$ are modified Bessel functions of first and second rank. 
The number of possible annihilation channels at the limit of zero velocity 
depends on the mass of the dark particle. In the aforementioned model, 
at tree level in perturbation theory the annihilation processes 
occur with exchanging a SM-higgs field, $h$ or a $\rho$ boson field. We consider 
a range of mass for DM where a pair of dark particles may annihilate 
through $s$-channel into a pair of SM fermions (quarks and leptons) and a 
pair of gauge bosons ($W^{+}W^{-},ZZ$) and also through $s$-, $t$- and $u$-channel into 
$hh$ or $\rho \rho$. We provide the necessary cross section formulas in the Appendix. 
The relevant Feynman diagrams for the dark matter annihilation processes 
are depicted in Fig.~(\ref{feynman}).

Employing the program LanHEP \cite{Semenov:LanHEP} we implement 
our model into the program CalcHEP \cite{Belyaev:CalcHEP} and calculate the 
annihilation cross sections and as a cross check on our implementation 
we find agreement with our analytical results given in the Appendix.
In the present article we analyze the relic density of DM by employing the program 
MicrOMEGAs \cite{Belanger:MICRO} which solves the Boltzmann equation numerically. 
MicrOMEGAs in turn uses the program CalcHEP to calculate 
all the relevant cross sections.
  
\subsection{Numerical Analysis: A first look}
\label{relic}
\begin{figure}
\begin{minipage}{0.40\textwidth}
\includegraphics[width=\textwidth,,angle =-90]{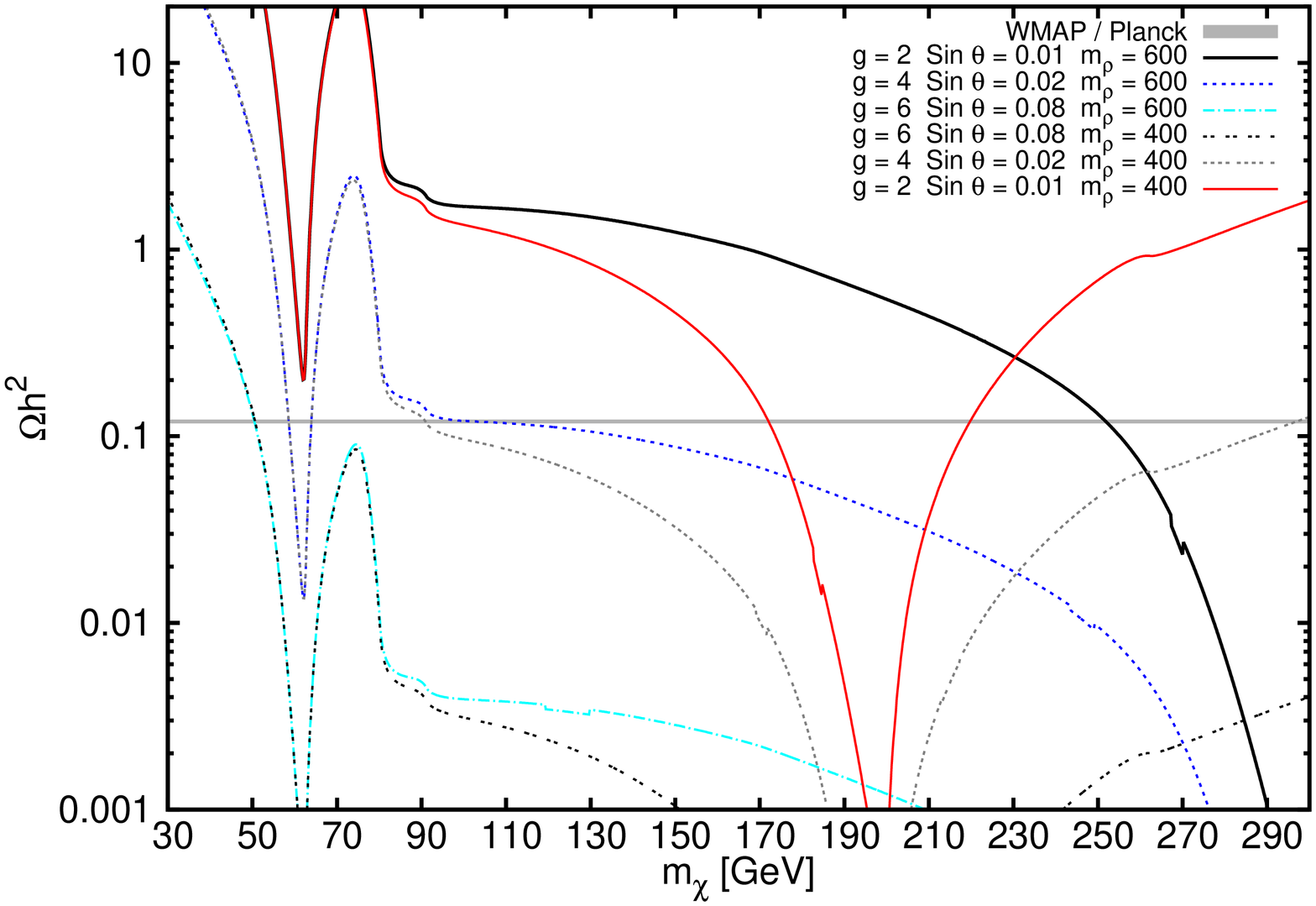}
\center{(a)} 
\end{minipage}
\hspace{1.8cm}
\begin{minipage}{0.40\textwidth}
\includegraphics[width=\textwidth,,angle =-90]{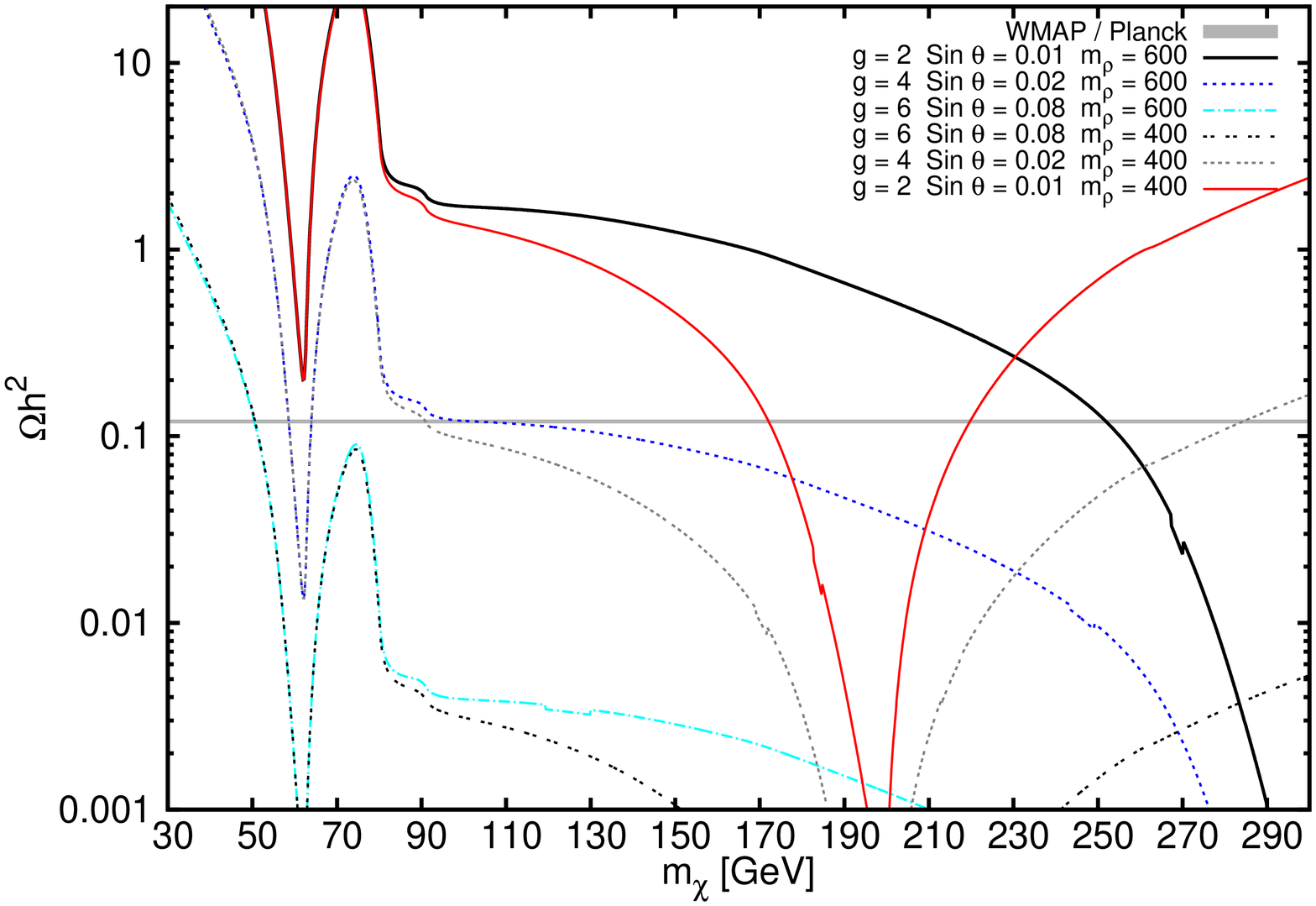}
\centerline{~~~~~~(b)}
\end{minipage}
\caption{Relic abundance as a function of DM mass. On the left plot $v_{\phi} = 600$ GeV
and $v_{\phi} = 1000$ GeV for the plot on the right. 
The horizontal band corresponds to $0.1172 < \Omega_{\text{DM}} h^2 < 0.1226$, 
which is a combined result from WMAP and Planck.}
\label{omega}
\end{figure}
We investigate here the viable region in the parameter space of the singlet fermionic model 
concerning the indirect detection of the fermionic dark matter along with implications
from invisible Higgs decay at LHC.       
Our analysis is performed with values for the quartic couplings 
which meet the constraints from vacuum stability and perturbativity condition. 

As a first numerical look we calculate the relic abundance as a function of DM mass
between 30 GeV and 300 GeV for two different values of $\rho$ 
boson mass, 400 GeV and 500 GeV. The constraint from invisible higgs decay is not
imposed here. 
The results depicted in Fig.~(\ref{omega}) 
for three choices of $g \sin \theta = 0.02, 0.08$ and $0.48$ show some correct  
characteristic features. One can see that the relic density drops fast 
for all set of parameters with $m_{\rho} = 400$ GeV, at DM mass close to 62 GeV 
and 200 GeV corresponding to the exchange of a SM-higgs and a singlet scalar, respectively. 
This sounds reasonable because the annihilation cross section get enhanced at resonance
regions and since $\Omega h^2 \propto (\sigma v)^{-1}$, a dip in the relic density should appear.    
Moreover, we expect some important effects on the relic density when $m_{\chi} \approx m_{W}$ and
$m_{\chi} \approx m_{Z}$ since at these masses two new channels now open up for DM to 
annihilate into. These effects show up in all the plots in Fig.~(\ref{omega}) at 
threshold values where the annihilation cross section increases and therefore 
make the relic density to decrease. One more additional study is done on the impact 
of the quartic couplings on the relic abundance. This can be done by adopting two 
distinct values for $v_{\phi}$, namely 600 GeV and 1000 GeV. We know already that 
only at large enough DM mass where two new channels $\bar \chi \chi \to hh$ 
and $\bar \chi \chi \to \rho \rho$ open, the size of $v_{\phi}$ becomes important
as the relations in Eq.~(\ref{couplings}) imply.
In Fig.~(\ref{omega}) the results when compared between left panel and right panel 
indicate that for smaller value of the quartic couplings
(corresponding to $v_{\phi} = 1000$ GeV) the relic density grows more significantly 
at DM mass close to $300$ GeV when $m_{\rho} = 400$ GeV as anticipated.      
\begin{figure}
\begin{center}
\includegraphics[width=.45\textwidth,angle =-90]{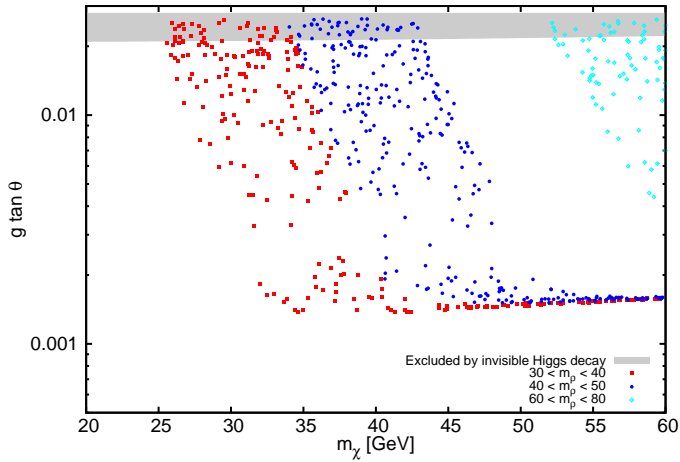}
\end{center}
\caption{The viable parameter set in the plane ($g \tan \theta,m_{\chi}$) with $m_{\chi}< m_{h}/2$ 
and $m_{\rho} < 80$ GeV. The gray band indicates excluded region
by invisible higgs decay.}
\label{scanro30-50}
\end{figure}

\subsection{Viable parameter space}  
\begin{figure}
\begin{minipage}{0.40\textwidth}
\includegraphics[width=\textwidth,,angle =-90]{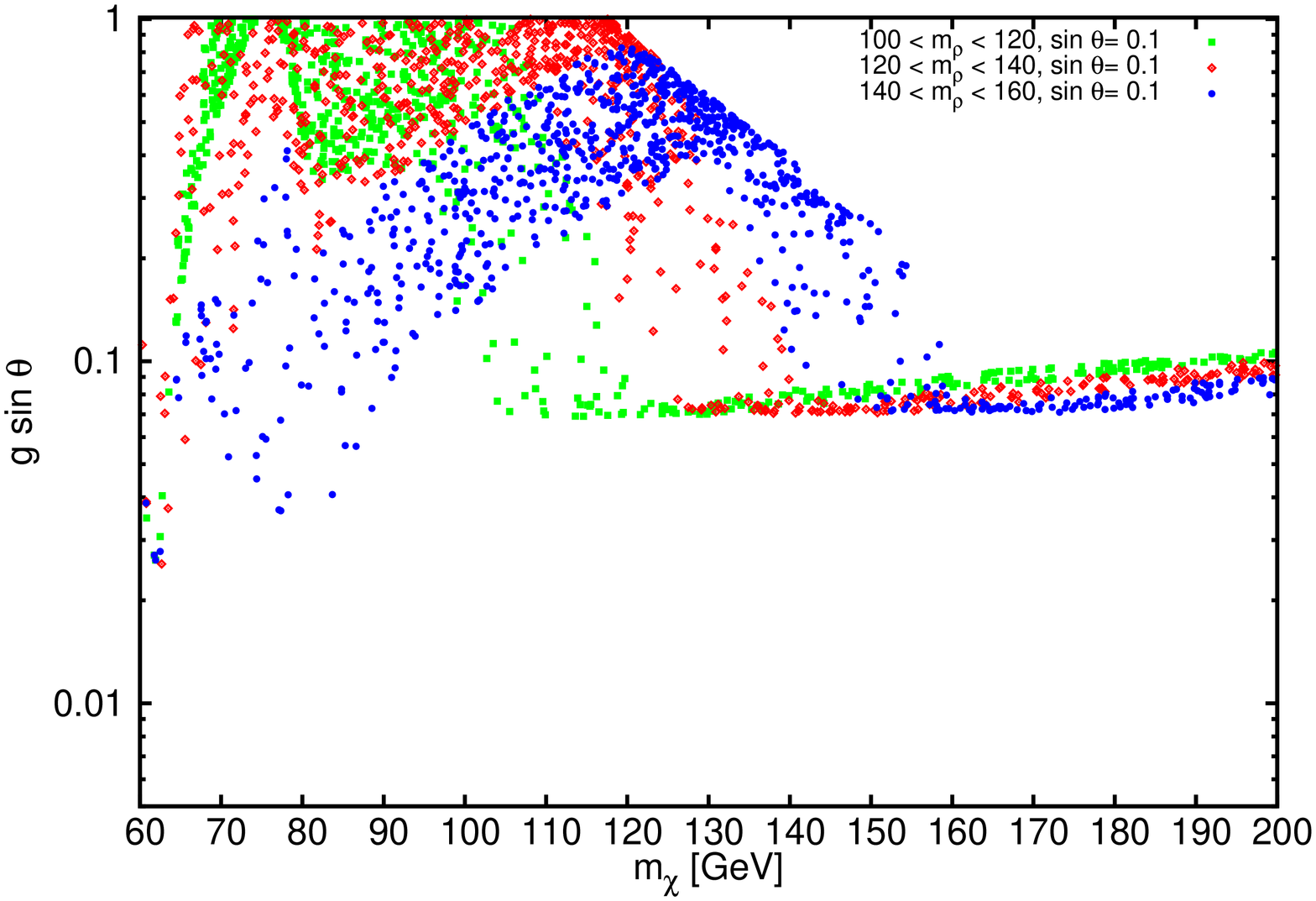}
\end{minipage}
\hspace{1.8cm}
\begin{minipage}{0.40\textwidth}
\includegraphics[width=\textwidth,,angle =-90]{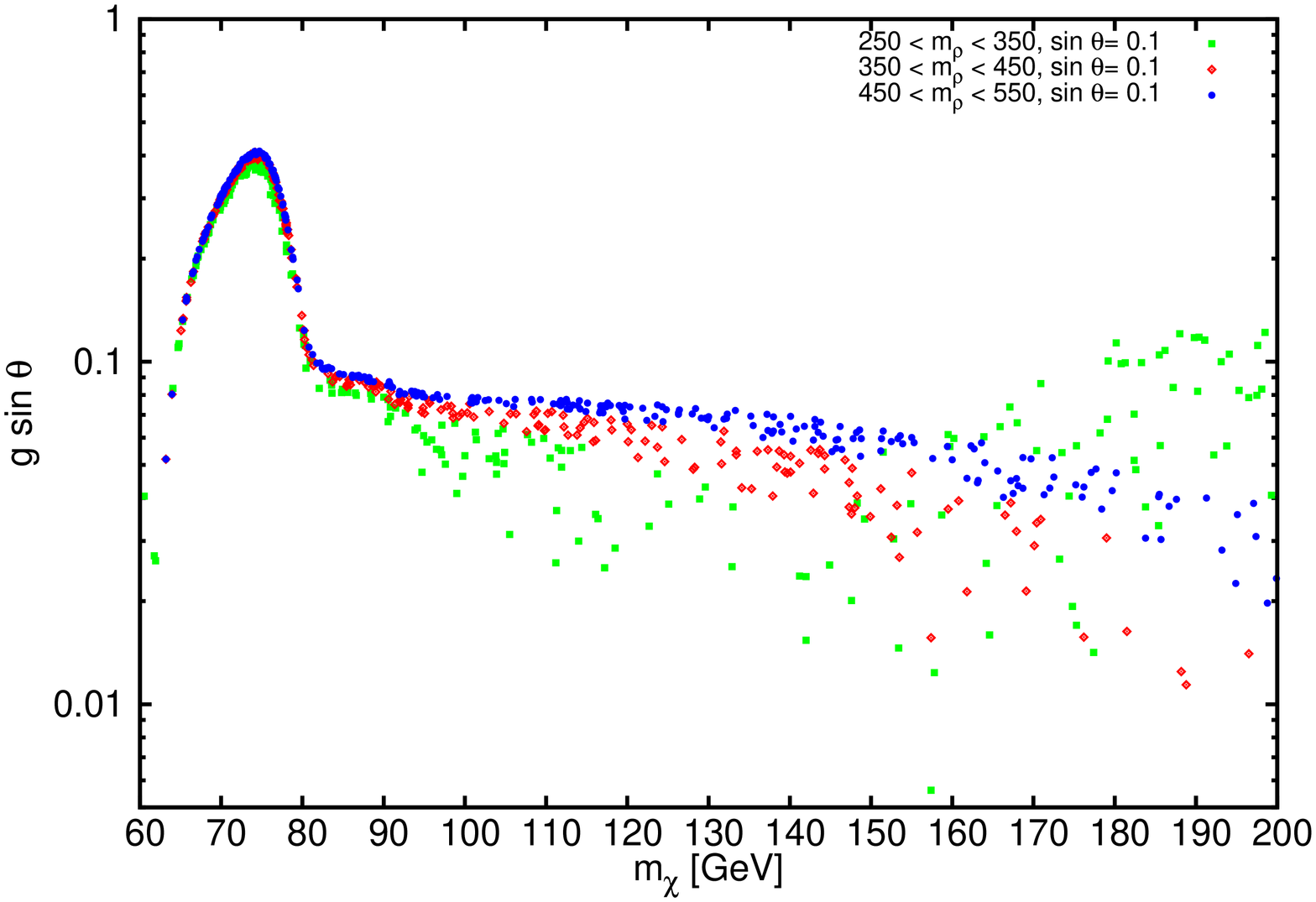}
\end{minipage}
\caption{The viable parameter set in the plane ($g \sin \theta,m_{\chi}$) with 60 GeV $<m_{\chi}< 200$ GeV 
and various intervals for $m_{\rho}$.}
\label{scangro100-550}
\end{figure}

\begin{figure}
\begin{center}
\includegraphics[width=.45\textwidth,angle =-90]{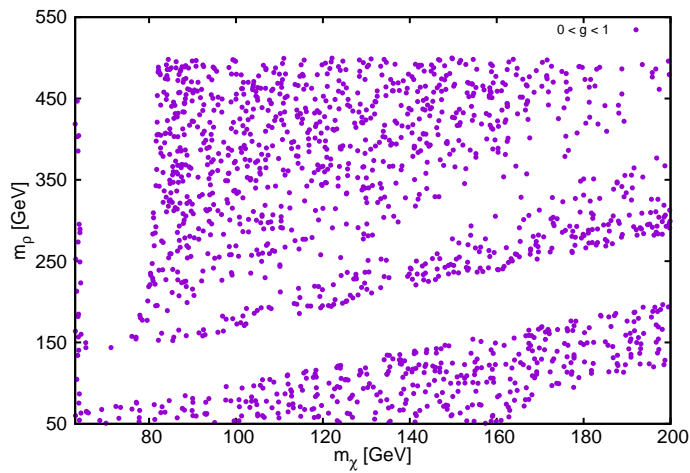}
\end{center}
\caption{The viable parameter set in the plane ($m_{\chi}$, $m_{\rho}$) with $0< g < 1$ 
and $\sin \theta = 0.1$ for masses 63 GeV $< m_{\chi} <$ 200 GeV and 50 GeV $< m_{\rho} <$ 500 GeV .}
\label{scandarkro}
\end{figure}
Taking into account the correct relic abundance of DM, we scan the parameter 
space over two ranges of the DM mass, namely $m_{\chi} < m_{h}/2$ 
and $m_{h}/2 < m_{\chi} < 200$ GeV. We have chosen this way because the invisible higgs 
decay put constrain on $g \sin \theta$ for $m_{\chi} < m_{h}/2$ but not on larger DM masses.

We first report on our results concerning the lower range mass, $m_{\chi} < m_{h}/2$ 
in Fig.~(\ref{scanro30-50}) where we take for the mixing angle 
such that $\sin \theta = 0.0026$ and generate random values for $g$ with $0 < g < 10$.    
The dominant DM annihilation channels are into final states $\bar b b$ and $\tau^+ \tau^-$
and in case we consider the region $m_{\rho} \lesssim m_{\chi}$, DM annihilation into $\rho \rho$ 
will take over at smaller values for $g \sin \theta$. As one important outcome, it is apparent from the 
figure that there is no allowed mass value for DM in the parameter space  when $m_{\rho}$ is far 
larger than $m_{\chi}$. This can be explained in terms of annihilation cross sections for two 
reactions $\bar \chi \chi \to \bar f f$ and $\bar \chi \chi \to \rho \rho$. When the latter reaction
becomes kinematically closed, in order for the total cross section to compensate the lack, it should 
pick up large values of $g \sin \theta$ which this may exceed allowed values determined by the 
invisible Higgs decay constraint.

Now, we look at the higher range for the DM mass, 60 GeV $< m_{\chi} <$ 200 GeV. Our results 
are provided by Fig.~(\ref{scangro100-550}) for two distinct interval for $m_{\rho}$, namely, 
100 GeV $< m_{\rho} <$ 160 GeV in the left panel and 250 GeV $< m_{\rho} <$ 550 GeV in the right panel. 
For both intervals we set $\sin \theta = 0.1$ and randomly generate $0 < g < 10$ and then single out 
the allowed region in the plane ($g \sin \theta, m_{\chi}$). We can notice from the figures that at $\rho$ 
boson masses less than 160 GeV, the coupling $g$ is allowed to pick out values larger than 
unity up to DM mass of about 150 GeV. For larger $\rho$ boson masses (250 GeV $< m_{\rho} < $550 GeV)
the coupling $g$ is allowed to exceed unity only at $m_{\chi} \lesssim$ 80 GeV and $m_{\chi} \gtrsim$ 170 GeV.     

Finally, in Fig.~(\ref{scandarkro}) we scan the viable region in ($m_{\chi},m_{\rho}$) 
space for reasonable values for the Yukawa coupling $g$, $0 < g < 1$ and a 
choice for the mixing angle such that $\sin \theta = 0.1$.     

\section{Direct detection}
\label{direct}
In this section we derive in detail the DM-nucleus elastic scattering cross section
for the model introduced above. 
The WIMP-nucleus elastic scattering cross section depends on the fundamental interaction 
of the WIMP-quark. The quark level interaction in our model  
occurs via t-channel by the Higgs exchange or the singlet scalar exchange, 
where at low momentum transfer the interaction is given by an effective 
four-fermi contact Lagrangian as
\ba
{\cal L_{\text{eff}}} = \alpha_{\text{q}} \bar \chi \gamma^5 \chi~ \bar q q \,,  
\ea 
with 
\ba
\alpha_{\text{q}} = (g \sin \theta \cos \theta) \frac{m_{\text{q}}}{v_{\text{H}}} (\frac{1}{m^{2}_{h}}-\frac{1}{m^{2}_{\rho}}).
\ea
We can now define the tree-level matrix element describing the scattering between 
the fermionic dark matter, $\chi$ and the individual nucleons $N$ (either proton $p$ or neutron $n$)
\ba
{\cal M_{\chi\text{N}} } = \sum_{\text{q}} \alpha_{\text{q}} \langle \chi_{f}| \bar \chi \gamma^5 \chi|\chi_{i} \rangle
 \langle N_{f}|\bar q q|N_{i} \rangle\,.
\ea
We cannot evaluate the nucleonic matrix element analytically because it is not known yet how
to connect the quark degrees of freedoms into the nucleonic ones through the non-petrubative 
mechanism of confinement.    
However, it is conventionally assumed that in the limit of vanishing momentum transfer, 
the nucleonic matrix element with the quark current is proportional to that 
with nucleon current \cite{Shifman:1978,Ellis:2008,Nihei:2004,Ellis:2000}  
\ba
\sum_{\text{q}} \alpha_{q} \langle N_{f}|\bar q q|N_{i} \rangle\  \to \alpha_{N} \langle N_{f}|\bar N N|N_{i} \rangle\ \,, 
\ea
where
\ba
\alpha_{N} = m_{N} \sum_{q = u,d,s} f^{N}_{Tq} \frac{\alpha_{q}}{m_{q}} 
+ \frac{2}{27} f^{N}_{Tg} \sum_{q = c,b,t}   \frac{\alpha_{q}}{m_{q}} \,.
\ea
The proportionality constants $f^{N}_{Tq}$ and $f^{N}_{Tg}$ incorporate the non-perturbative
physics of strong interaction at low energy and $m_{N}$ represents the nucleon mass.
To proceed we shall follow closely the discussions in \cite{Dienes:2013,DelNobile:2013,Anand:2013}. 
We can now construct the matrix element for the dark matter-nucleus scattering
in the non-relativistic limit as
\ba
{\cal M_{\chi \text{T}}} =  \alpha_{N} \langle N_{f}|\bar N N|N_{i} \rangle\ 
  \langle \chi_{f}| \bar \chi \gamma^5 \chi|\chi_{i} \rangle 
\approx 4 \alpha_{N} m_{N} (\frac{m_{T}}{m_{N}}) (\xi^{s^\prime}_{N})^\dagger \xi^{s}_{N} 
\langle \chi_{f}|\vec q.\vec S_{\chi} | \chi_{i}  \rangle \,,
\ea  
where the matrix element $\langle \chi_{f}|\vec q.\vec S_{\chi} | \chi_{i}  \rangle$ 
contains the DM-spin operator in which $\vec q$ is the momentum transferred to 
the nucleus and $\xi^{s}_{N}$ is the two-component spinor corresponding to the fermion N 
with spin $s$. The extra factor $\frac{m_{T}}{m_{N}}$ is inserted due to the different 
normalization between the target nucleus with mass $m_{T}$ and the nucleon with mass $m_{N}$.  
We therefore obtain the corresponding squared matrix element averaged over initial spin states 
and summed over the final states as
\ba  
\langle|{\cal M_{\chi \text{T}}}|^2\rangle = \frac{s_{\chi}(s_{\chi}+1)}{3(2j_{T}+1)} 
16 \alpha^{2}_{N} m^{2}_{T} \vec q\,^2
\ea
The differential cross section for DM-nucleus scattering in the non-relativistic limit 
\cite{DelNobile:2013} reads
\ba
\frac{d\sigma_{T}(v,{\vec q}\,^{2})}{d {\vec q}\,^2} = 
\frac{\langle|{\cal M_{\chi \text{T}}}|^2\rangle}{64\pi m^{2}_{\chi}m^{2}_{T} v^2}\,.
\ea
We can then calculate the total cross section as
\ba
\sigma (v) = \int^{4v^2 \mu^{2}_{T}}_{0} \frac{d\sigma_{T}(v,{\vec q}\,^{2})}{d {\vec q}\,^2} d\vec q\,^2 \,,
\ea
where $\mu_{\chi T}$ is the reduced mass of the DM-nucleus system. 
We arrive finally at our expression for the spin-independent (SI) total cross section for DM-nucleus scattering
\ba
\sigma_{\text{SI}} (v) =  \frac{[Z \alpha_{p}+(A-Z)\alpha_{n}]^2}{2j_{T}+1} \frac{v^2\mu^{4}_{T}}{2\pi m^{2}_{\chi}} \,,
\ea
where the DM-nucleus relative velocity $v \sim {\cal O}(10^{-3})$. We note that the 
DM bilinear matrix element results in a velocity suppression in the cross section 
of WIMP-nucleus in the no-relativistic limit. Our numerical probe over 
the full parameter space shows that DM scattering rate in our model 
is far below the minimum bands imposed on the scattering rate 
by the current results from LUX and XENON100, so that the dark matter particle 
can evade direct detection. 
Thus, as it was known from earlier works, we expect no 
constraints on the parameter space of our model from direct detection of DM.    

\section{Gamma-ray emission from DM self-annihilation}

The evidence for the gamma-ray emission from a small region 
centered on the Galactic Center originating from annihilating dark matter
was pointed out firstly in \cite{Goodenough:2009} based on data from 
{\it Fermi} Gamma-Ray Space Telescope. 
Further studies with confirmation on this finding can be found 
in \cite{Hooper:2010mq,Boyarsky:2010dr,Hooper:2011ti,Abazajian:2012pn,Gordon:2013vta,Abazajian:2014fta}.    
Other sources, in particular, unresolved
millisecond pulsars are also considered to explain the observed 
anomalous gamma emission from the Inner Galaxy \cite{Hooper:2011ti,Abazajian:2012pn,Abazajian:2010zy}.
However, recent studies relying on an estimated population of millisecond pulsars in the Inner
Galaxy suggest that millisecond pulsars 
make up only a small portion ($< 5\%$ ) 
of the total observed gamma excess, see discussions in \cite{Cholis:2014lta,Cholis:2014noa}.  
In the following we assume that the observed gamma-rays are produced as a 
result of DM annihilation in the Inner Galaxy.
We shall then discuss in this section the gamma-ray emission from dark matter 
self-annihilation in the singlet fermionic dark matter model. 
Here, we restrict our attention to DM mass below $W^\pm$ 
and $Z$ threshold, thus dark matter annihilation takes place 
only with SM fermions in the final states ($\bar \chi \chi \to \bar f f$) 
via SM-higgs exchange or singlet scalar exchange.  
The flux of gamma-rays at Earth produced by annihilating dark matter 
located in the central region of the Milky Way is 
\ba
\frac{d^2\Phi}{dE_{\gamma}d\Omega} = \frac{1}{16\pi}\frac{\langle \sigma v \rangle_{ann}}{m^{2}_{DM}}
\frac{dN_{\gamma}}{dE_{\gamma}} \int^{\infty}_{0} dr \rho^{2}(r^\prime,\theta) \,,
\ea   
where the distance from the annihilation point to the earth denoted by $r$ and
$r^\prime$ is given in terms of the angle between the line of sight and 
the center of the galaxy as $r^{\prime} = \sqrt{r^{2}_{\odot}+r^2-2r_{\odot}r\cos \theta}$ 
with $r_{\odot} = 8.5$ kpc. The photon flux depends upon two dynamical quantities,
the annihilation cross section times the relative velocity, $\langle \sigma v \rangle_{\text{ann}}$
and the photon energy spectrum generated per self-annihilation into a 
fermion pair, $dN_{\gamma}/dE_{\gamma}$.
\label{gama-ray}
\begin{figure}
\begin{center}
\includegraphics[width=.50\textwidth,angle =-90]{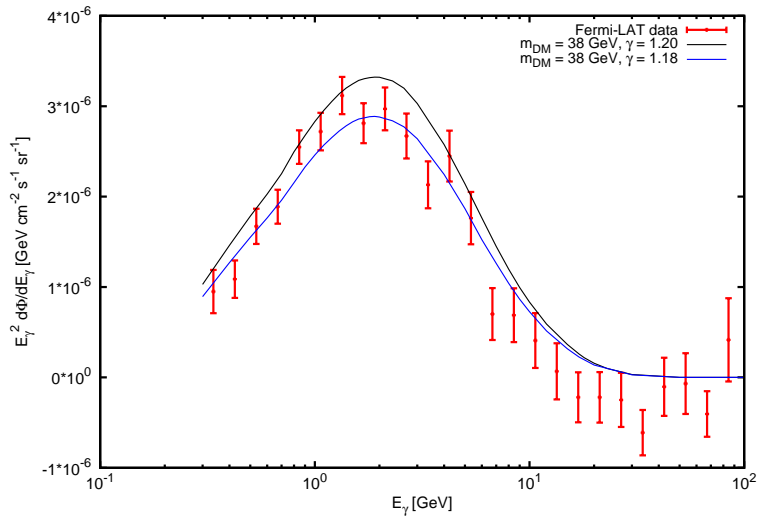}
\end{center}
\caption{The flux of gamma-ray excess as data points are shown 
at $5 \degree$ from the Galactic Center \cite{Daylan:2014rsa}. Gamma-ray spectra from 
dark matter annihilation into fermion pair at $m_{\text{DM}}$= 38 GeV are compared
for two different values of the inner slope, $\gamma$.} 
\label{gamaexcess}
\end{figure}    
It is assumed that the dark matter distribution is approximately spherical and thus 
we can give the dark matter density as a function of the distance from the Galactic Center, $r$. 
Throughout our study we use the DM density characterized as
\ba
\rho(r) = \rho_{\odot} (\frac{r_{\odot}}{r})^\gamma \Big(\frac{r_{c}+r_{\odot}}{r_{c}+r} \Big)^{3-\gamma}\,,
\ea 
where $\gamma = 1$ is the standard NFW value for the inner slope. 
The scale radius chosen as $r_{c} = 20$ kpc and $\rho_{\odot} = 0.3~\text{GeV/cm}^3$ is the local
dark matter density at $8.5$ kpc from the Galactic Center.
We employ the package MicrOMEGAs to calculate the gamma-ray spectrum. 
Since the astrophysical parameters involved in our computation for the gamma-ray flux
are given with uncertainties we do not limit ourself to the region in the parameter space 
which precisely meet the constraints from observed relic abundance and invisible higgs decay width. 
Our results for the gamma-ray excess is presented 
in Fig.~(\ref{gamaexcess}) for $m_{\text{DM}} = 38$ GeV as an example, with 
two values for the inner slope, $\gamma =$ 1.18 and 1.20. The singlet scalar mass 
is chosen as $m_{\rho} = 76$ GeV (this is the resonance mass and 
enhance the cross section significantly). It turns out that the dominant annihilation channels are 
into $\bar b b$ quark pair ($\sim 94\%$) and $\tau^+ \tau^-$ ($\sim 6 \%$) with the total 
annihilation cross section $\langle \sigma v  \rangle_{\text{ann}} \sim 1.7 \times 10^{-26}$ cm$^3$s$^{-1}$
consistent with the value demanded by the thermal relic.  
We compare our results with the Fermi-LAT data for the extended gamma-ray 
excess extracted from \cite{Daylan:2014rsa}.
As it is evident from the plots in Fig.~(\ref{gamaexcess}), the gamma-ray 
flux with $\gamma =$ 1.18 gives a better fit to the Fermi-LAT data.

\section{Conclusions}
\label{con}
In this work we considered a fermionic dark matter model in which dark matter particle 
communicates with the SM particles via a CP-violated interaction term. 
It is known that in this model dark matter evades direct detection because the 
DM-nucleus elastic scattering cross section is velocity suppressed.  
The invisible higgs decay width is computed within the model and constraint
on the product $g \tan \theta$ is found.   
We then calculated the DM annihilation cross sections analytically and confirmed  
our results by applying the package CalcHEP. The restricted region in the 
parameter space is then found in consistent with the observed DM relic density.

We realized that the fermionic model discussed above indicates a robust connection between invisible 
Higgs decay and indirect detection signals. 
In fact we emphasize that in the parameter space can be found a CP-violated coupling for DM with 
natural magnitude which meets the constraint from the anticipated invisible 
Higgs decay width and satisfy the indirect detection restrictions provided 
by Planck and WMAP observations.   

Furthermore, assuming that the origin of Fermi-LAT gamma-ray excess is due to the 
WIMP dark matter annihilation in the Galactic Center, we note that a fermionic dark
matter with $m_{\text{DM}} \sim 38$ GeV can account for this excess.   

Therefore, it is crucial that in the light of null result from current direct 
detection experiments, within a model with Yukawa pseudo scalar DM coupling
it is possible to understand indirect detection signals.

\section{Acknowledgments}
I would like to thank Hossein Ghorbani for useful discussions.
\label{Ack}

\section{Appendix: Annihilation cross sections}
\label{Apen}
We obtain the annihilation cross section of a DM pair into a pair of SM fermions as
\ba
\label{ff}
\sigma_{\text{ann}} v_{\text{rel}} (\bar \chi \chi \to \bar f f) = 
\frac{g^2 \sin^2 2\theta}{64 \pi} 
\Big[  \frac{1}{(s-m^{2}_{h})^2+m^{2}_{h}\Gamma^{2}_{h}}
\nonumber\\&&\hspace{-8.5cm}
+\frac{1}{(s-m^{2}_{\rho})^2+m^{2}_{\rho}\Gamma^{2}_{\rho}} 
-\frac{2(s-m^{2}_{h})(s-m^{2}_{\rho})+2m_{h}m_{\rho}\Gamma_{h}\Gamma_{\rho}}
{((s-m^{2}_{h})^2+m^{2}_{h}\Gamma^{2}_{h})((s-m^{2}_{\rho})^2+m^{2}_{\rho}\Gamma^{2}_{\rho})} \Big] \times
\nonumber\\&&\hspace{-8.5cm}
 \Big(  N_{c} \times 2 s (\frac{m_{f}}{v_{H}})^2 (1-\frac{4m^{2}_{f}}{s})^{\frac{3}{2}} \Big) \,,
\ea
where $N_{c}$ is the number of color charge. The dominant contributions belong to 
the heavier final states $b \bar b$ and $t \bar t$.  
The total cross section into a pair of gauge bosons in unitary gauge is given by 
\ba
\label{ww}
\sigma_{\text{ann}} v_{\text{rel}} (\bar \chi \chi \to W^+ W^- , ZZ) = 
\frac{g^2 \sin^2 2\theta}{64 \pi} 
\Big[  \frac{1}{(s-m^{2}_{h})^2+m^{2}_{h}\Gamma^{2}_{h}} 
\nonumber\\&&\hspace{-10cm}
+\frac{1}{(s-m^{2}_{\rho})^2+m^{2}_{\rho}\Gamma^{2}_{\rho}} 
-\frac{2(s-m^{2}_{h})(s-m^{2}_{\rho})+2m_{h}m_{\rho}\Gamma_{h}\Gamma_{\rho}}
{((s-m^{2}_{h})^2+m^{2}_{h}\Gamma^{2}_{h})((s-m^{2}_{\rho})^2+m^{2}_{\rho}\Gamma^{2}_{\rho})} \Big] \times
\nonumber\\&&\hspace{-10cm}
 \Big[ (\frac{m^{2}_{W}}{v_{\phi}})^2(2+\frac{(s-2m^{2}_{W})^2}{4m^{4}_{W}}) (1-\frac{4m^{2}_{W}}{s})^{\frac{1}{2}}
\nonumber\\&&\hspace{-10cm}
+\frac{1}{2}(\frac{m^{2}_{Z}}{v_{H}})^2(2+\frac{(s-2m^{2}_{Z})^2}{4m^{4}_{Z}}) (1-\frac{4m^{2}_{Z}}{s})^{\frac{1}{2}}
     \Big] \,.
\ea
And finally we get the following result for the annihilation scattering into two higgs bosons as
\ba
\label{hh}
\sigma_{\text{ann}} v_{\text{rel}} (\bar \chi \chi \to h h) = 
\frac{g^2}{32 \pi} (1-\frac{4m^{2}_{h}}{s})^{\frac{1}{2}}
\Big[  \frac{a^2 \sin^2 \theta}{(s-m^{2}_{h})^2+m^{2}_{h}\Gamma^{2}_{h}} 
\nonumber\\&&\hspace{-10cm}
+\frac{b^2 \cos^2 \theta}{(s-m^{2}_{\rho})^2+m^{2}_{\rho}\Gamma^{2}_{\rho}} 
+\frac{ a b \sin 2\theta [(s-m^{2}_{h})(s-m^{2}_{\rho})+m_{h}m_{\rho}\Gamma_{h}\Gamma_{\rho}]}
{((s-m^{2}_{h})^2+m^{2}_{h}\Gamma^{2}_{h})((s-m^{2}_{\rho})^2+m^{2}_{\rho}\Gamma^{2}_{\rho})} \Big] 
\nonumber\\&&\hspace{-10cm}
+ \frac{g^4 \sin^4 \theta}{16 \pi s} (1-\frac{4m^{2}_{h}}{s})^{\frac{1}{2}} 
\Big[s(m^{2}_{\chi}-t)+m^{2}_{\chi}m^{2}_{h}-(m^{2}_{\chi}+m^{2}_{h}-t)^2 \Big]  \times
\nonumber\\&&\hspace{-9.85cm}
\Big(\frac{1}{t-m^{2}_{\chi}}+\frac{1}{u-m^{2}_{\chi}}\Big)^2 \,,
\ea
with 
\ba
a = \sin^3 \theta \lambda v_{\phi} +6 \cos^3 \theta \lambda_{H} v_{H}
         +6 \sin^2 \theta \cos \theta \lambda_{1} v_{H} + 6 \cos^2 \theta \sin \theta \lambda_{1} v_{\phi} 
\nonumber\\\nonumber\\&&\hspace{-12.8cm}
b = \cos \theta \sin^2 \theta \lambda v_{\phi} -6 \cos^2 \theta \sin \theta \lambda_{H} v_{H}
-6 \sin^3 \theta \lambda_{1} v_{H} + 4 \sin \theta \lambda_{1} v_{H}
\nonumber\\\nonumber\\&&\hspace{-12cm}
-6 \cos \theta \sin^2 \theta \lambda_{1} v_{\phi}+ 2 \cos \theta \lambda_{1} v_{\phi} \,,      
\ea
and annihilation cross section into two $\rho$ boson is
\ba
\label{rr}
\sigma_{\text{ann}} v_{\text{rel}} (\bar \chi \chi \to \rho \rho) = 
\frac{g^2}{32 \pi} (1-\frac{4m^{2}_{\rho}}{s})^{\frac{1}{2}}
\Big[  \frac{c^2 \sin^2 \theta}{(s-m^{2}_{h})^2+m^{2}_{h}\Gamma^{2}_{h}} 
\nonumber\\&&\hspace{-10cm}
+\frac{d^2 \cos^2 \theta}{(s-m^{2}_{\rho})^2+m^{2}_{\rho}\Gamma^{2}_{\rho}} 
+\frac{c~d \sin 2\theta [ (s-m^{2}_{h})(s-m^{2}_{\rho})+m_{h}m_{\rho}\Gamma_{h}\Gamma_{\rho}]}
{((s-m^{2}_{h})^2+m^{2}_{h}\Gamma^{2}_{h})((s-m^{2}_{\rho})^2+m^{2}_{\rho}\Gamma^{2}_{\rho})} \Big] 
\nonumber\\&&\hspace{-10cm}
+ \frac{g^4 \cos^4 \theta}{16 \pi s} (1-\frac{4m^{2}_{\rho}}{s})^{\frac{1}{2}} 
\Big[s(m^{2}_{\chi}-t)+m^{2}_{\chi}m^{2}_{\rho}-(m^{2}_{\chi}+m^{2}_{\rho}-t)^2 \Big]  \times
\nonumber\\&&\hspace{-9.85cm}
\Big(\frac{1}{t-m^{2}_{\chi}}+\frac{1}{u-m^{2}_{\chi}}\Big)^2 \,,
\ea
with 
\ba
c = \lambda v_{\phi} \cos^3 \theta  -6 \lambda_{H} v_{H} \sin^3 \theta 
         -6 \lambda_{1} v_{H} \cos^2 \theta \sin \theta  + 6 \lambda_{1} v_{\phi} \sin^2 \theta \cos \theta  
\nonumber\\\nonumber\\&&\hspace{-12.8cm}
d = \lambda v_{\phi}  \sin \theta \cos^2 \theta  +6 \lambda_{H} v_{H} \sin^2 \theta \cos \theta 
-6 \lambda_{1} v_{H} \sin^2 \theta \cos \theta  + 2 \lambda_{1} v_{H} \cos \theta 
\nonumber\\\nonumber\\&&\hspace{-12cm}
+6 \lambda_{1} v_{\phi} \sin^3 \theta  - 4 \lambda_{1} v_{\phi} \sin \theta  \,.      
\ea

\end{document}